# Haptic Manipulation of Microspheres Using Optical Tweezers Under the Guidance of Artificial Force Fields

Ibrahim Bukusoglu[*], Cagatay Basdogan[*†], Alper Kiraz[+], Adnan Kurt[+]

(*) College of Engineering, Koc University, Istanbul Turkey, 34450
(+) College of Arts and Sciences, Koc University, Istanbul Turkey, 34450

## Abstract

Using optical tweezers and a haptic device, microspheres having diameters ranging from 3 to 4 μm (floating in a fluid solution) are manipulated in order to form patterns of coupled optical microresonators by assembling the spheres via chemical binding. For this purpose, biotin-coated microspheres trapped by a laser beam are steered and chemically attached to an immobilized streptavidin-coated sphere (i.e. anchor sphere) one by one using an XYZ piezo scanner controlled by a haptic device. The positions of all spheres in the scene are detected using a CCD camera and a collision-free path for each manipulated sphere is generated using the potential field approach. The forces acting on the manipulated particle due to the viscosity of the fluid and the artificial potential field are scaled and displayed to the user through the haptic device for better guidance and control during steering. In addition, a virtual fixture is implemented such that the desired angle of approach and strength are achieved during the binding phase. Our experimental studies in virtual and real environments with 8 human subjects show that haptic feedback significantly improves the user performance by reducing the task completion time, the number of undesired collisions during steering, and the positional errors during binding. To our knowledge, this is the first time that a haptic device is coupled with OT to guide the user during an optical manipulation task involving steering and assembly of microspheres to construct a coupled microresonator.

[†] Corresponding Author: cbasdogan@ku.edu.tr

# 1 Introduction

Optical Tweezers (OT) constitute a single beam optical trap which can be used to immobilize or transport objects in three dimensions. The first optical trap was built by Ashkin in 1970. He showed that small dielectric particles can be accelerated by the radiation pressure of a laser beam and trapped by two counter propagating beams. In 1986, Ashkin, Dziedzic, Bjorlholm and Chu succeeded in trapping particles using a single, highly focused laser beam known as OT. The basics of optical trapping using OT are quite straightforward for both objects smaller and larger than the wavelength of light. A laser beam is focused by a high numerical aperture microscope objective to a spot in the plane of a specimen. This spot creates an optical trap which is able to hold a small particle at its focus. The electric dipole moment in response to the light's electric field creates a force towards the focus of the light beam on the objects. For small objects, the trapping force is given by the gradient of the electric field. For large objects, it can be visualized as the reaction to the net momentum change of photons passing through the particle. This force moves the particle when the laser beam is moved or immobilizes the particle when the beam is stationary and the fluid solution is translated with respect to the beam using an XYZ piezo scanner. The majority of OT makes use of conventional $TEM_{00}$ Gaussian beams to trap particles. Laguerre Gaussian beams are also used in the OT to trap and manipulate objects that are optically reflective or absorptive. Laguerre Gaussian beams have a well defined orbital angular momentum that can also rotate trapped objects. Also a zero or higher order Bessel beam is used in OT to rotate and move multiple particles apart from each other and around obstacles. A single-trap standard OT has one or at most two laser beams. On the other hand, multiple-trap OT (i.e. Holographic OT) are made possible by sharing a single beam in temporal or spatial space using acousto-optic deflectors or galvanometer driven mirrors (Castelino, Satyanarayana, and Sitti, 2005).

OT are suitable for manipulating objects with length scales ranging from tens of nanometers to micrometers and exerting forces on the objects ranging from femtonewtons to nanonewtons (Rohrbach and Stelzer, 2002). Due to these features and its noninvasive nature (because no physical contact is present), OT have found applications in many fields. In particular, they have been used for investigating biological systems since the scale of the interaction forces are highly small (Grier, 2003). For example, OT have been used to probe the viscoelastic properties of single biopolymers (such as DNA), cell membranes, aggregated protein fibres (such as actin), gels of such fibres in the cytoskeleton, and composite structures (such as chromatin and chromosomes). They have also been used to characterize the forces exerted by molecular motors such as myosin, kinesin, processive enzymes and ribosomes. It has been proposed that OT can also be used to construct patterns by assembling micro particles. These patterns can be used later as building blocks for constructing micro devices as suggested, but not implemented, in Castelino et al. (2005).

Manipulating micro-scale particles using OT under the guidance of visual feedback only is highly challenging. When an external force is applied to a particle trapped by a laser beam, the particle is displaced from the focus of the beam in a manner similar to an object that is attached to a mechanical spring and exposed to an external force. In addition, drag force acting on the particle (note that particles are floating in a fluid solution) during steering makes the manipulation more difficult. For example, the particle may easily escape from the trap due to the drag force if the manipulation velocity is too high. Moreover, if two particles are intended to be attached to each other via chemical binding as in some biological applications, one must not only avoid undesired collisions with other particles during the steering of each particle, but also align them precisely during binding. Collision-free steering using visual feedback only is challenging if there are several particles floating in the solution. Similarly, checking if binding has occurred is not straightforward. For example, Bryant et al. (2003)

binds two DNA coated microspheres to each other using a pipette and visual feedback in order to measure the mechanical properties of DNA. He measures the forces acting on the trapped sphere in X and Y directions using a photo-detector first and then displays their magnitude using graphical bars on the computer screen to see if binding has occurred. Force and position control is also necessary if a micro or nano scale object is to be attached to or inserted into a biological entity. For example, Pauzauskie et al. (2006) use OT to insert a fluorescent nanowire into a cell for tracking its movements in the body. They manipulate the wire and bring it close to a human cervical cancer cell first and then push the wire carefully for several seconds to penetrate inside the cell without damaging it. As it is obvious from these examples, in many applications of OT, collision-free steering, accurate positioning, and force-controlled manipulation are important for the successful execution of the task.

To this end, haptic feedback appears to be a natural choice (Arai et al., 1999; Bukusoglu, Basdogan, Kiraz, and Kurt, 2006). Using an OT and a haptic device, we show that coupled microsphere optical resonators in various patterns can be formed more accurately and efficiently by steering and binding biotin-coated microspheres to streptavidin-coated ones. During steering and assembly, we display the drag force and artificial guidance forces acting on the manipulated sphere to the user through the haptic device. Under the guidance of these forces, a biotin-coated sphere (trapped by a laser) is steered along a collision free path and chemically bound to a streptavidin-coated sphere along a pre-calculated line specifying the angle of contact. A virtual fixture pulls the trapped sphere towards this line when it is sufficiently close to the anchor sphere. Virtual fixtures and artificial force fields have been shown to improve user performance and learning in telemanipulation tasks in real world and training tasks simulated in virtual environments (Rosenberg, 1993; Payandeh and Stanisic, 2002; Bettini et al. 2002). They are also effective in robot programming by demonstration (Aleotti, Caselli, and Reggiani, 2005).

The remainder of the paper is organized as follows: Section 2 introduces the concept of coupled microsphere optical resonators and the current problems in this area. In Section 3, we propose an alternative method to construct these resonators by manipulating microspheres using a haptic device and OT. We also provide a review of the earlier work in the same section. Section 4 describes the components of our experimental setup to achieve this goal. In Section 5, we discuss the calibration of the setup. Section 6 describes the haptic assistance modes, including an artificial potential field used for steering, and a virtual fixture used for binding the manipulated spheres. In Section 7, we discuss the novel application of the proposed methods to the construction of coupled microsphere resonators in detail. Section 8 presents the results of two experiments conducted separately in virtual and real environments to investigate the role of haptic guidance in steering and binding microspheres to form a coupled microsphere resonator. Finally, the discussion of the results and the conclusions are given in Section 9.

**2 Microsphere Optical Resonators**

In a single sphere optical resonator, light confined in the spherical volume with an index of refraction greater than that of the surrounding medium exhibits resonances called Whispering Galley Modes (WGMs) (Vahala, 2003). WGMs can be observed by analyzing the emission spectrum of the fluorescent molecules doped inside a microsphere. As shown in Figure 1c, high intensity peaks observed in the emission spectrum indicate the WGMs. An important parameter in quantifying these peaks is the quality factor Q, which is defined as $\omega/\Delta\omega$, where $\omega$ is the resonance frequency and $\Delta\omega$ is the bandwidth at that resonance frequency. An ideal resonator would confine light indefinitely (without loss) leading to an infinite Q. However, infinite Q cannot be reached because of the absorption, reflection, and scattering losses (Eversole, Lin, and Campillo, 1995).

Microsphere optical resonators are systems where high Q factors can be achieved in small mode volumes. This makes them very attractive for applications in atom optics, quantum electrodynamics, optical communication, and biology. Some applications of the resonators in these fields include microlasers, narrow filters, optical switching, displacement measurements, Raman sources, studies of nonlinear optical effects and ultrafine sensing. The research in this area has recently focused on coupled microsphere resonators. Coupled resonators can be formed by assembling microspheres in different patterns. Möller, Woggon, and Artemyev (2004, 2005, and 2006) showed the changes in the emission spectrum of exactly size matched microspheres doped with CdSe quantum dots arranged in different orientations. They also demonstrated waveguiding of WGMs using coupled microspheres. In these studies, coupled resonators are formed by a self-assembly process. The spheres are first mixed in a solution of methanol and then left to dry to form random patterns. The self-assembly process has two major drawbacks: First, the nature of the process prevents accurate positioning and alignment of the patterns. Accurate positioning is important as it determines the frequency of the coupled optical modes in optical resonators. Furthermore, the waveguiding is determined by the accurate alignment of the microspheres. Second, the variety of patterns that can be formed by a self-assembly process is limited, and complex patterns cannot be constructed. This limits the number of resonators that can be constructed for different applications.

**3 Our Approach**

As an alternative to the self-assembly process, we propose to use haptic feedback in optical manipulation to construct coupled optical resonators in a controlled manner. For this purpose, we developed a setup to steer and assemble microspheres using OT and a haptic device. We demonstrate that this approach helps the user manipulate the spheres more effectively and efficiently in assembling different patterns. To our knowledge, this is the first time that a haptic device is coupled with OT to guide the user during an optical manipulation task involving transportation and assembly of microspheres to construct a coupled optical resonator. In comparison to our study, Lee, Lyons, and LeBrun (2003) propose a model to simulate force interactions between a particle and a laser beam for haptic manipulation of micro particles, but their work is purely theoretical and haptic forces are used to keep the particle in the trap only and not for steering and forming an assembly. In our approach, in addition to displaying trapping forces to the user via a haptic device (we, in fact, display the drag force during steering, which is balanced by the trapping force), an artificial force field is used to help the user steer the trapped particle to the desired location more efficiently while avoiding collisions with the other particles and also to bind particles to each other with high precision. This level of precision is important for coupled microsphere resonators and coupled resonator optical waveguides. Similar to our approach in principle, Arai et al. (1999, 2000) also use a haptic device for optical manipulation. They manipulate biological samples indirectly using "microtools" including a micro bead, a micro basket, and a micro capsule. These tools are trapped by the laser beam and positioned by the haptic device connected to a XY positioning stage. They are primarily used to manipulate and isolate targeted biological objects such as Escherichia coli and yeast cell. Again, the haptic feedback is used to convey trapping forces to the user and not for guidance during steering and assembly. Moreover, a user study showing the benefits of the haptic feedback is missing. Hence, for example, if the task performance is better or the user can learn to execute a manipulation task more quickly with haptic feedback has not been investigated.

    We should also point out that there are a number of studies involving haptic devices and manipulation of micro particles using a probe (Bettini, Lang, Okamura, and Hager, 2002) and also nano particles using the tip of scanning probe of an Atomic Force Microscope (AFM) (Taylor II, 1994, Sitti and Hashimoto, 2003, Li, Xi, Yu, and Fung, 2003). The probes

used in these studies are always in physical contact with the manipulated object. The manipulated particle may easily stick to the probe due to adhesive forces or the probe may easily slip over the particle. Moreover, there is a lower limit for the size of the objects being visualized and manipulated in micro manipulation. In nano manipulation, surface scanning and manipulation are done sequentially by the same probe, which is highly time consuming (i.e. you need to scan and determine the location of particles first to manipulate them later). Moreover, due to several environmental uncertainties (e.g. ground vibrations and changes in humidity and temperature affect the measurement of probe and particle locations), finding the location of particles precisely and achieving robust manipulation are highly challenging in nano manipulation. In comparison, manipulations with OT are non-invasive. This is particularly important when working with biological samples, since an AFM probe may easily contaminate or damage the manipulated object under investigation during scanning or manipulation. Moreover, OT can be used to manipulate objects in nano-scale. On the other hand, uncontrolled laser power in OT may also damage the biological samples. Moreover, it may also be difficult to determine the exact location of particles floating in a fluid due to their Brownian motion. In addition, if the fluid around the assembled object must be dried after the manipulations, as in our case, some undesired results may be observed due to the surface tension of the fluid. For example, we observed that the constructed patterns of optical resonators may sometimes break or unwanted bindings of the surrounding particles to the constructed pattern may occur during the drying stage.

## 4 Set-Up

The setup for the proposed manipulation system is shown in Figure 2. The major components of this system are OT and a haptic interface. For the OT, the beam of a continuous wave green laser (Crysta Laser CRL-GCL-025-L, $\lambda = 532$ nm) with an output power of 25 mW is sent through a 6x magnifying telescope into an inverted microscope (Nikon TE 2000-U). After being reflected off a dichroic mirror (Chroma Filters Q570LP), the laser beam is focused on the sample by a high numerical aperture (NA = 1.4, 60x) microscope objective. This beam is used to trap microspheres for manipulation. A CCD camera is used to capture the images of the spheres for visual feedback to the user and to calculate their location in the image. By using an intermediate magnification module, a total magnification of 90x is achieved. A red pass filter (Chroma filters HQ610/75) is used to filter out stray laser light. In our system, the location of the laser beam is fixed and the movements of a sphere trapped by the beam is controlled by a three-dimensional piezoelectric scanner (Tritor 102 SG from Piezosystem Jena Inc.) working in the closed-loop control (scanning resolution is 2 nm). The movements of the scanner are commanded by the user via a haptic device (Omni from Sensable Technologies Inc.) The displacements of the haptic stylus are scaled and suitable voltage values are sent to the scanner to control its movements on the sample plane. The forces acting on the trapped particle during the manipulations are scaled up and conveyed to the user through the same haptic device.

The components of the setup are synchronized via a computer program written in C++ language. The flowchart of the program is given in Figure 3. First, a snapshot of the manipulation environment is captured via the CCD camera. Then a threshold is applied to the captured image and center position and radius of each microsphere in the image are determined using a contour finding algorithm. Using this information, a virtual model of the scene is constructed to provide the user with the bird's eye view of the manipulation environment during the execution of the task. Since the CCD camera is stationary and not moving with the scanner as the spheres move, the user may easily loose the "big picture" during the steering of a sphere without the bird's eye view. Following the construction of virtual model, the user enters the necessary inputs according to the nature of manipulation task to be performed. For example, a target location for the manipulated sphere is entered for a steering task. If the task involves forming an assembly as in our application, then the location of the anchor sphere and the binding angles are specified. Based on the input parameters and the location of the spheres in the scene, the computer program automatically generates an artificial force field to provide haptic guidance to the user during the execution of the task.

## 5 Calibration

The setup must be calibrated in order to find the limits of the manipulation. Since the manipulation process is carried out in a fluid solution, drag forces acting on the trapped particle may exceed the forces applied by the laser beam if the particle is manipulated with a high velocity. A laser beam exerts gradient and scattering forces on the trapped object. Small objects having sizes comparable to laser wavelength develop an electric dipole moment in response to the light's electric field. This leads to a gradient force which attracts the particle to the beam focus with a magnitude proportional to the intensity gradient of the laser beam. Larger objects such as the microspheres used in our experiments act as lenses, causing to a change in the momentum of the incident photons. This results in an effective force which draws the particle towards the higher flux of photons near the focus (Grier, 2003). For the inverted geometry depicted in Figure 4a, the particle reaches equilibrium along the axis of the laser beam (z-axis) by the gradient, scattering, and the gravitational forces. As long as the particle is at the center, the intensity of the light around the particle is symmetric and hence the gradient forces cancel each other resulting in a forward force (scattering force) which is balanced by the weight of the particle (Figure 4b). However, when the centers of the trapped particle and the laser beam do not coincide with each other, the gradient force pulls the particle to the center of the beam (Figure 4c). During the manipulations, the gradient force exerted by the laser beam is balanced by the drag force acting on the trapped particle. If the manipulation velocity is high, the particle may escape from the trap. In order to avoid this situation, the maximum achievable velocity in our setup is determined prior to the manipulations. This process is also known as the calibration.

There are various techniques available to calibrate OT. The calibration approach that is most appropriate to use with a CCD camera is the escape force method (Visscher, Gross, and Block, 1996). In this method, a trapped particle is moved away from the cover slip such that the distance between the particle and the cover slip is much larger than the radius of the particle. Then, the piezo scanner is commanded to move in increasingly higher velocities until the particle escapes from the trap while the displacement of the trapped particle from the center of the beam is measured and recorded (Figure 5). The drag force applied on the particle is calculated using the Stokes law, $F_{DRAG} = 6\pi\eta rv$, where $\eta$ is the viscosity of the fluid, v is the velocity of the scanner and r is the radius of the particle. We found that the particle escapes from the trap when the velocity exceeds 100 μ/sec. A correction factor is typically applied to the escape velocity according to the Faxen's law for the spherical objects manipulated close to a cover slip (Svoboda and Block, 1996). The correction factor for our application is calculated as 2.36. After the correction is applied, the maximum velocity for controlled manipulation of a particle in our optical trap is reduced to 42 μm/sec (100/2.36).

**6 Optical Manipulation with Haptic Feedback**

The haptic device in our setup controls the movements of the XYZ scanner to manipulate the particles and also provides force feedback to the user during the manipulations. The haptic device and OT are both commanded by the same computer; hence time delays during teleoperation are not significant. As the user manipulates the trapped particle via the haptic device, the movements of the particle are tracked from the camera image as discussed in Section 3. The positional scaling between the movements of the haptic stylus and the XYZ scanner was adjusted such that one millimeter movement of the haptic stylus was equivalent to 3 pixels of movement of the trapped particle in the image (the size of one pixel is 200 nm, hence the scale factor in positional movements is 1667). The forces acting on the trapped particle during the manipulations are scaled up and displayed to the user via the haptic device. In our implementation, the maximum total force displayed to the user is saturated to 2N. Arai et al. (1999) use the same teleoperation approach with different scaling factors in position and force.

In order to assemble microspheres to construct a coupled microresonator, two types of manipulation are necessary: 1) the spheres must be transported to the target locations (*steering*) and then 2) must be attached to each other by chemical means (*binding*). We use a potential field approach to help user steer a particle while avoiding the collisions with the others. During the binding, a virtual fixture is used to position the particles precisely and also to apply controlled forces to the user to make the binding process easier. Hence, the forces conveyed to the user during the manipulations can be classified into three groups: the forces due to a) the artificial potential field (i.e. path planning forces), b) the virtual fixture, and c) viscous drag of the fluid.

**a) Path Planning Forces:** For individual steering of particles, we have implemented a path planning algorithm based on the potential field approach. This is a well-known approach in robotics used for motion planning (Khatib, 1986; Choset et al., 2005). Potential field approach involves the modeling of the position to be reached as an attractive pole and obstacles as repulsive surfaces. In our application, an artificial potential field is constructed from the components associated with the goal location where the trapped particle is to be steered, $U_{goal}$, and obstacles (untrapped particles) along the steering path, $U_{obstacle}$ (see Figure 6). Then, the net potential acting on the particle is

$$U(q) = U_{goal}(q) + \sum U_{obstacle}(q) \qquad (1)$$

where, q is the location of the trapped particle in the potential field. Typically, $U_{goal}$, is defined as a parabolic attractor in the form of

$$U_{goal}(q) = \frac{1}{2}\alpha\, dist(q, q_{goal}) \qquad (2)$$

where, $\alpha$ is the constant gain.

The repulsive force of an obstacle is typically modeled as a potential barrier that arises to infinity as the particle approaches to the obstacle. It is also desired that the repulsive potential does not affect the steering of the particle when it is sufficiently away from the obstacle. The repulsive field of an obstacle is defined as

$$U_{obstacle}(q) = \begin{cases} \frac{1}{\beta} g(q)((3r^2 - g(q)^2)) & \text{if } g(q) \leq r \\ 0 & \text{if } g(q) > r \end{cases} \qquad (3)$$

where,

        g(q) : Euclidian distance between trapped particle and obstacle

        r    : radius of influence

        $\beta$    : constant gain

The radius of influence was taken as r = 1.2*(radius of the obstacle + radius of the trapped particle being steered). The coefficient "1.2" takes into account the drift in obstacle position due to the Brownian motion.

The net force acting on the particle is the negative gradient of the total potential $F = -\nabla U(q)$. As a result of the repulsive forces, a possible collision with the other particles along the steering path is avoided. These collisions may result in the escape of the manipulated particle from the trap or an undesired binding of the manipulated particle to the collided one. The attractive forces, on the other hand, give an intuition to the user about how he/she should push the particle in order to reach the target location for binding. These two forces together create a tunneling effect and guide the user along a collision-free path.

**b) Drag Force:** As discussed earlier, a viscous drag force acts on the trapped particle while it is being steered. The drag force acts opposite to the direction of movement and is balanced by the trapping force of the laser beam. When the drag force exceeds the trapping force, the particle escapes from the trap. In order to prevent this happening, a drag force proportional to the velocity of particle is applied to the user during the manipulation. We prefer to display drag force to the user through the haptic device since it is easier to calculate (i.e. the trapping force is proportional with the amount of shift between the centers of the trapped particle and the laser beam, which is more difficult to measure precisely during steering). The damping coefficient in the drag force function is set such that the user does not exceed the maximum allowable steering velocity of 42 μm/sec (corresponds to a maximum speed of 70 mm/sec for the haptic stylus) and the total force displayed to the user does not exceed the maximum allowable force of 2 N. As a result, the escape of the particle from the trap is eliminated and a more efficient manipulation is achieved.

**c) Virtual Fixtures:** The term virtual fixture refers to a software implemented guidance that helps the user perform a task by limiting his/her movements into restricted regions and/or influencing its movement along a desired path (Rosenberg, 1993). The virtual fixtures can be thought of as a ruler or a stencil (Abott and Okamura, 2003). By the help of a ruler or stencil, a person can draw lines and shapes faster and more precisely than the ones drawn by free hand.

Similarly, a haptic device can be programmed to apply forces to the user in a virtual environment to perform a task more efficiently and precisely. In comparison to the physical constraints, the type and the number of virtual constraints that could be programmed is unlimited.

Virtual fixtures offer an excellent balance between automated operation and direct human control. They can be programmed to help the operator carry out a structured task faster and more precisely or they can act as safety elements preventing the manipulators from entering dangerous or undesired regions. For example, studies on telemanipulation systems show that user performance on a given task can increase as much as 70% with the introduction of virtual fixtures (Rosenberg, 1993). Other applications of virtual fixtures include robotic assisted surgery, robot programming by demonstration, and training (Abbot and Okamura, 2003; Aleotti et al., 2005).

In our application, we use a virtual fixture to provide haptic guidance to the user during binding such that he/she stays on the correct approaching direction. When the distance between the manipulated sphere and the one fixed on the cover slip (the anchor sphere) is less than a threshold value (i.e. less than 1.2 times the sum of radii of the manipulated and the fixed spheres), the virtual fixture becomes active and pulls the user towards a straight line along the desired binding angle passing through the center of the anchor sphere. The user is free to move along the line, but when he/she moves out of the line, an attractive spring force with a stiffness coefficient of $k = 0.2$ N/mm and proportional to the perpendicular distance between the line and the position of the microsphere is applied to him/her by the haptic device (see Figure 7). This helps the user to accomplish the binding task more easily and accurately. Moreover, a repulsive field is defined around the anchor sphere to prevent the user from applying excessive forces during binding which results in escape of the binding particle from the trap.

## 7 Assembling Microspheres

In order to assemble microspheres in different patterns, it is necessary to bind microspheres to each other. Currently, only simple patterns can be formed by self-assembly. We suggest that more complex patterns can be constructed using an OT, a haptic device, and the proposed haptic manipulation and guidance techniques. Binding of microparticles to each other have been accomplished using different mechanisms in the past. The examples of these mechanisms include antibody-antigen and protein-receptor bindings and binding achieved using DNA molecules (Mirkin, Letsinger, Mucic, and Storhoff, 1996; Alivisatos et al., 1996) and hydrophobic interactions (Onoe, Matsumoto, and Shimoyama, 2004). Since the particles in all of these approaches are attached to each other via self-assembly, control on the pattern formation is very limited.

In our approach, binding of the microspheres to each other is also accomplished via the protein-receptor mechanism, but in a controlled manner. Particles coated with streptavidin molecules (protein) and biotin molecules (receptor) are chemically bound to each other. The streptavidin-biotin couple is commonly used for binding particles through self-assembly due to its extreme stability over a wide range of temperature and pH. Streptavidin-biotin binding has been proposed in Castelino et al. (2005) as a method for constructing micro-structures, but not implemented. In our application, streptavidin coated microspheres (i.e. anchor spheres) are immobilized on the coverslip first. Then, a biotin coated microsphere is trapped by the OT and steered it to bring close to a streptavidin coated microsphere with the help of path planning forces. Finally, these two particles are chemically bound to each other precisely with the help of a virtual fixture. The binding of the biotin to the suitable docking sides on the streptavidin occurs rapidly as a result of the strong affinity between them (Figure 8a). Since the desired angle of binding is set by the user in advance, different geometric patterns can be constructed (Figure 8b).

# 8 Experimental Study

In order to investigate the role of the haptic guidance in optical manipulation, we have designed two separate experiments. Eight human subjects participated to experiments. In both experiments, there were two experimental conditions depending on the type of sensory feedback displayed to the subjects:

(1) Only visual feedback was displayed to the subjects (V)
(2) Visual and haptic feedback was displayed together (V+H)

## 8.1 Experiment I:

**Goal**: The goal of the first experiment was to analyze the task performance of the subjects during particle steering with and without the help of haptic feedback. The task was to construct a coupled microsphere resonator made of 4 spheres by steering and binding three spheres to an anchor sphere individually. This experiment was conducted in a virtual world because it was not possible to display equal experimental conditions to all subjects in real world settings. In order to assemble microspheres using our approach, first a droplet of solution containing streptavidin and biotin coated particles must be put on a cover slip and then biotin coated spheres must be trapped, steered, and chemically attached to streptavidin coated ones. It is not possible to control the number of particles in the droplet and their distribution on the cover slip. Hence, it was not possible to display the same manipulation environment and the experimental conditions to all subjects during the steering phase of the task. For this reason, the simulation experiment was designed to test primarily the task performance of users during the steering phase of the task.

**Stimuli:** Subjects were asked to steer and assemble 4 dark blue spheres around an anchor sphere (light blue) in virtual environments as shown in Figure 9 (the size of the spheres is assumed to be 4 μm). They were asked to pick each of the four dark blue spheres using their haptic device in the given order and steer it to the goal position displayed on the screen (pink sphere) for binding to the anchor sphere while avoiding collisions with the green spheres. A collision free-path connecting each of the dark blue spheres to the goal position was calculated in advance using the artificial field approach given in Section 6. Forces due to the artificial force field and viscous drag were displayed to the subject during the steering process. The attractive and repulsive gain constants were set to $\alpha = 0.007$ and $\beta = 1200$, respectively in Equations 2 and 3. All distances in those equations were calculated in millimeters.

**Design:** A matrix approach was followed to eliminate the effect of the order of experimental conditions on the performance of the subjects. The subjects were divided into two groups (four in each group) and repeated the same task 20 times under each sensory condition. The experiments were performed in two sets with one week time interval between the sets. In the first set, the first group received ten trials with condition (2) (V+H) and then ten trials with condition (1) (V) while the second group received ten trials with condition (1) and ten trials with condition (2). After one week, the first group received ten trials with condition (1) and then ten trials with condition (2) while the second group received ten trials with condition (2) and ten trials with condition (1) (Table 1).

**Procedure:** Before the experiments, subjects were given a handout explaining the goal of the experiment and containing step by step instructions about what they should do to finish the task successfully. The subject was instructed to move the haptic cursor (shown as a small red sphere in the screen) over the first dark blue sphere and press the "S" key on the keyboard to select (i.e. trap) it. Once the sphere was trapped, the haptic device was virtually coupled to it and a transparent pink sphere appeared on the screen, showing the goal position. Then the subject steered the trapped sphere by moving the haptic stylus to the goal position while avoiding collisions with other spheres in the scene (green spheres). When the task was accomplished, the subject dropped the manipulated sphere by pressing the "D" key on the keyboard. The same procedure was repeated for the other 3 dark blue spheres shown in Figure 9.

**Results:** A total of 6 different measures were defined to evaluate the performance of the subjects under two different experimental conditions. One of the measures of the task completion time is the total time that passes between the picking and dropping of the four spheres. The rest of the measures can be classified into two groups. The first group of performance measures is related to the steering process and includes contact duration and maximum penetration during steering. The contact duration during steering is the total time that the manipulated sphere is in contact with the obstacles (green spheres) during its transfer to the goal position. The maximum penetration during steering is the maximum distance the manipulated sphere penetrates into the obstacles during steering. The second group of performance measures is related to the binding process and includes contact duration, maximum penetration, and maximum angular deviation during binding. The contact duration during binding is the total time that the grabbed sphere is in contact with the anchor sphere (light blue sphere). The maximum penetration during binding is the maximum distance the manipulated sphere penetrates into the anchor sphere during binding. The maximum angular deviation during binding is the maximum angle between the line connecting the center of the anchor sphere and the goal position and the one connecting the centers of the anchor and the manipulated spheres while they are in contact. These 6 performance measures were calculated after each trial and the mean values of 20 trials were used to compare the performance of the subject under the experimental conditions of visual feedback only (V) and visual and haptic feedback together (V+H). The results are shown in Figure 10. The figure shows that the performance under (V+H) is significantly better than that of (V) for all performance measures ($p < 0.05$). The haptic feedback improves both the speed and accuracy of the steering and binding processes. These results agree with the results of the earlier teleoperation experiments. We further investigated the effect of haptic feedback on task learning. For this purpose, the average task completion times of the subjects with and without haptic feedback are plotted against the number of trials. As shown in Figure 11, the task completion time under the condition (V) does not change much with the number of trials (indicating no learning) while it decays rapidly under the condition (V+H) indicating that the task is being learned. An exponential curve in the form of $y = ae^{-bx} + c$ is fitted to the experimental data of the (V+H) condition ($R^2 = 0.996$). Using the equation, it is estimated that the curve reaches to a steady state (i.e. task learning is accomplished) at $14^{th}$ trial with 5 % relative error.

## 8.2. Experiment II:

**Goal:** A second experiment was conducted using the OT setup to investigate the subjects' performance during the binding process with and without the guidance of haptic feedback. The goal of the second experiment was to assemble 3 spheres to form a coupled resonator by chemically binding two mobile biotin-coated particles to the left and right sides of a streptavidin-coated particle fixed on the cover slip. Subjects were asked to form a horizontal line by carefully aligning the spheres. Accurate alignment and positioning of spheres in a coupled microsphere resonator is important for obtaining the desired emission spectrum.

**Stimuli:** The experiment was conducted in real world setting using the OT setup integrated with a haptic device. We only focused on the performance of subjects during the binding process since the steering environment (i.e. the location and distribution of the particles) varied from trial to trial and subject to subject. The radius of each obstacle and the trapped particle were determined from camera image as discussed in Section 4. The attractive and repulsive gain constants were set to $\alpha = 0.007$ and $\beta = 1200$, respectively in Equations 2 and 3. All distances in those equations were calculated in terms of image pixels and then converted to millimeters. In addition to the image window, a simulation window showing the virtual models of the particles was displayed to the subjects. During the experiments, the subject steered the trapped particle to the binding site while tracking its movements in the image and the simulation windows simultaneously. The binding site was displayed as a red sphere in the simulation window to guide the subject.

**Procedure:** To construct a resonator made of 3 spheres aligned on a horizontal line, the subject trapped a biotin particle, steered it to the suitable binding site, and then attached it to a streptavidin-coated microsphere by light pushing. It was possible to differentiate a fluorescent streptavidin coated particle from a non-fluorescent biotin coated one by simply directing the laser beam on it (a biotin-coated sphere does not emit light and floats in the solution). To trap a biotin coated sphere, the subject manipulated the haptic stylus and controlled the XYZ scanner and brought the particle under the laser beam. Then, the laser filter was turned off by the subject by pressing the "W" key on the keyboard to trap the particle. The filter was initially active to prevent the particle escaping from the trap accidentally. If a biotin particle was trapped successfully (if not, the subject pressed the "E" key to turn on the filter again and searched for another particle), then the subject initialized the visual and haptic simulation by pressing the "S" key on the keyboard. The subject attached a biotin-coated sphere to a streptavidin-coated one by mating them first and then carefully applying a pushing force. Subjects were asked to test if the binding occurred by lightly shaking the biotin particle using the stylus of their haptic device. If the subject was convinced that the binding had occurred, he/she finalized the task by pressing the ON/OFF switch on the haptic stylus.

**Sample Preparation:** The biotin-coated non-fluorescent microspheres with diameters ranging from 3 to 3.9 μm were purchased from Kisker-Biotec Inc. (ID: PC-B-03). The dragon green fluorescent microspheres with a nominal diameter of 3.69 μm (ID: FS05F) were purchased from Bangs Laboratories Inc and then coated with streptavidin molecules (purchased from Prozyme Inc.) in the bio-technology laboratory of our school. We preferred to use fluorescent microspheres for two reasons: first, the spheres must be fluorescent in order to be used as optical resonators. Second, a fluorescent particle emits light when a laser with an appropriate wave length is directed on it and hence this feature can be used to differentiate streptavidin-coated particles from the biotin-coated ones which are non-fluorescent. The fluorescent microspheres were coated with streptavidin using a passive adsorption technique. First, 100 μl of solution containing microspheres was centrifuged at 6500 rpm for 5 minutes. At the end of the centrifugation process, the supernatant was removed and discarded. Then, the microspheres were resuspended in water and vortexed for mixing. This washing process was repeated twice. But in the second run, the microspheres were resuspended in PBS with Ph = 7.4 instead of water. After the last washing, the appropriate amount of purified streptavidin (1 μl) was added to the microsphere solution and mixed gently for 2 hours at room temperature. Then, the solution was left over night at 4 ºC. The process was finalized after the solution was washed and resuspended in PBS again as before. The biotin-coated microspheres were suspended in a different storage buffer then the streptavidin-coated particles. In order to prevent the biotin coated particles sticking to the cover slip, we added 1% volume of BSA into the PBS solution.

To prepare the manipulation environment for our experiments, we first placed the coverslip on the XYZ piezo-scanner. Then, we dropped 25 μl of solution containing streptavidin-coated microspheres on the coverslip and waited until the microspheres stick to the coverslip surface. Finally, 25 μl of solution containing biotin-coated microspheres was added.

**Design:** The same group of 8 subjects participated in the second experiment. The subjects were asked to construct exactly 10 assemblies (i.e. resonators) under two sensory conditions (V and V+H). The order of the sensory feedback displayed to the subjects was random, but same for all subjects. The number of trials performed by each subject to construct the required number of 10 resonators depended on his/her success in the task. If the manipulated sphere escaped from the optical trap during an experimental trial, that trial was counted as "failed".

**Results:** As shown in Table 2, the number of trials performed by the subjects to construct 10 resonators under the condition (V) is more than the one performed under the condition (V+H). The reason for the "unsuccessful" trials under the condition (V) is the escape of the particle from the trap during binding process due to excessive penetration to the anchor sphere. Two additional measures were defined to evaluate the performance of the subjects under each sensory condition: 1) angular deviation during binding on the plane of coverslip and 2) elevation difference between the attached and anchor spheres along the direction perpendicular to the plane of coverslip (see Figure 12). The angular deviations of the biotin-coated spheres chemically bound to the streptavidin-coated sphere from the left and right on the plane of cover slip ($\alpha_L$ and $\alpha_R$) were measured from the camera image at the end of each successful trial. Then, the average of the left and right angular deviations was calculated for 10 trials. Similarly, the elevation difference between each attached sphere and the anchor sphere ($Z_L$ and $Z_R$) were measured using the focus of the camera and the movements of the scanner along the axis perpendicular to the plane of cover slip. The results (Figure 13) show that the average angular deviation and the elevation difference are significantly higher for the sensory condition (V) ($p < 0.05$).

**9 Discussion and Conclusion**

Our personal communications with expert physicists revealed that techniques for precise manipulation and alignment of spheres to construct different patterns of microresonators with desired optical properties are insufficient and relies on visual information only. The techniques available today for this purpose include self assembly, self assembly on lithographically patterned surfaces, and individual manipulation of microspheres using tapered fiber probes, but none of them is sufficient for building complex structures with high precision. In self-assembly, the patterns are formed randomly with no control on the geometry. In invasive manipulation approaches using fiber probes, electrostatic forces affect the positioning accuracy of the particles. However, the performance of these structures as resonators depend on the size and positional accuracy of the assembled spheres which can have a profound, but not well studied effect on optical transport (Astratov, Franchak, and Ashili, 2004). In fact, even the concept of coupled microsphere resonators is new and only recently Möller et al. (2004) demonstrated the existence of optical coupling for a microresonator made of 3 spheres aligned on a line (Note that the size of the microspheres, coupled via self assembly, is identical within a deviation of less than 0.1%). The effect of variations in the pattern geometry on the emission spectrum of the resonator is still an open research problem and there are only very few published studies on this subject. Guo, Quan and Pau (2005) performed computer simulations on single silicon nitride microdisks with diameters of 2 μm and 10 μm positioned at a certain distance away from an optical waveguide. They observed that an optimal gap existed between the microdisk and the waveguide for maximum energy coupling. The optimum gap was found to be a strong function of the wavelength of the resonant mode. In a 10 μm diameter microdisk, for a resonant mode at 608 nm, normalized stored energy was calculated to drop by 64% when the gap changed from 250 nm to 200 nm. Guo, Quan, and Pau (2005) also found that the Q-factor of the resonator increases exponentially with increasing gap and saturates as the gap approaches the optical wavelength. These results point at the importance of the alignment and positioning when assembling coupled microsphere resonators.

In this study, we showed that displaying guidance forces through a haptic device improves the task learning and the performance of an operator significantly in constructing a coupled microresonator by steering and assembling microspheres. To assemble the microspheres in our application, we used biotin-streptavidin binding. This type of chemical binding takes place rapidly. Due to this reason, collisions of the manipulated particle with the other particles during steering may lead to unintended binding if the manipulations are performed with visual feedback only. To prevent this, artificial repulsive force fields were constructed around the particles along the manipulation path and these forces were displayed to the subject through the haptic device. These repulsive forces combined with the attractive forces to pull the manipulated particle to the goal point by generating a tunneling effect. Our experiments conducted in virtual environments showed that haptic guidance of artificial force fields improved the user performance and learning significantly. A supporting conclusion was also achieved by Kuang, Payandeh, Zheng, Henigman, and MacKenzie (2004) in a study investigating the role of virtual fixtures in training. They showed that significant learning and training transfer occur with force field guidance of virtual fixtures measured by performance time and path length. Another benefit of the haptic feedback in our application is in the binding phase. The repulsive field around the selected anchor sphere prevented the subject from applying excessive forces to it while trying to bind a manipulated particle. Moreover, virtual fixtures were used for the precise attachment of the manipulated particles to the anchor particle in order to form an assembly with a desired pattern. The virtual fixture exerted forces on the manipulated particle to push it towards a pre-defined line of contact for successful binding. The results of our second experimental study showed that the binding achieved using visual feedback only causes elevation differences as large as 3 μm (Mean = 0.93 ± 1.07 μm) between the attached and anchor spheres along the axis perpendicular to the coverslip, which is a significant deviation compared to the diameter of the assembled spheres (~ 4 μm). Moreover, we also observed that the angular deviation on the plane of coverslip during binding could be as large as 20° degrees (Mean = 6.1 ± 4.9°) without haptic feedback. These values are highly significant if we consider the fact that even the coupling efficiency and the Q factor of a single sphere resonator change drastically with a few hundred nanometer changes in the gap between the sphere and the waveguide (Guo, Quan and Pau, 2005). The results of our study showed that the desired patterns for microresonators could be constructed more precisely using haptic feedback (0.05 ± 0.15 μm and 1.3 ± 1.7°). Moreover, the failure rate (i.e. the number of times the manipulated particle escapes from the trap) was much lower when haptic feedback exists in the system (see Table 2). In particular, the elevation difference is highly significant for the condition (V). We observed that subjects pushed the trapped sphere uncontrollably towards the anchor sphere for binding when there was only visual feedback. Due to the contact interactions, the trapped sphere slid below or above the anchor sphere creating an elevation difference along the axis perpendicular to the coverslip. Since the CCD camera displays the top view of the spheres located on the coverslip only, adjusting the position of the manipulated sphere with respect to the anchor sphere during binding was difficult. When haptic feedback exists, the artificial forces due to the repulsive field of the anchor sphere prevented the subject to apply excessive forces during binding. Moreover, the artificial forces due to the virtual fixture pushed the trapped sphere towards the desired binding direction and kept it there. The virtual fixture shown in Figure 7 is in the form a 3D cylinder and pushes the particle towards the center line. Since the user does not have a direct visual feedback along the direction of the laser beam (Z axis), the haptic feedback helps him/her adjust the Z position.

We should point out that the steering and binding processes discussed in this paper could be fully automated. However, better image processing, computer recognition, and control techniques are necessary 1) to identify the particle type (streptavidin versus biotin), 2) to test if the particle is mobile or not, 3) to automate the trapping of a biotin particle, 4) to continuously track the positions of the mobile biotin particles in the scene, and 5) to test if binding has occurred or not. Alternatively, a shared control architecture can be developed such that the automated steering and human-operated binding could be combined together (Sheridan, 1992; O'Malley, Gupta, Gen, & Li, 2006). However, the presence of a human operator in the steering task has an additional benefit. It is known that the potential field approaches suffer from the local minima problem (Koren and Borenstein, 1991) and the steering task may be terminated by the computer without reaching the goal if automated methods are used. However, when a human operator is present in the loop, he/she can push the particle to escape from the local minima and guide it towards the right direction by intuition.

**This study demonstrates that haptic feedback significantly improves the user performance and task learning in optical manipulation. Considering the fact that OT are used by many research groups in the world to manipulate micro and nano scale objects with applications in various fields, this study may stimulate the others to integrate haptic devices into their applications for more precise and controlled manipulation and may open the door to new applications. For example, a laser beam controlled by a haptic device can be used to dissect or manipulate cells or chromosomes. It is also possible to construct a holographic OT and use a multi-fingered haptic device to perform more complex manipulations. For example, two beams, controlled by a haptic device each, can be used to manipulate micro structures to form an assembly or**

to construct molecular motors. Similarly, the same set-up can be used to manipulate a florescent wire to insert into a cancerous cell in order to track its movements in a blood stream.

**Figures:**

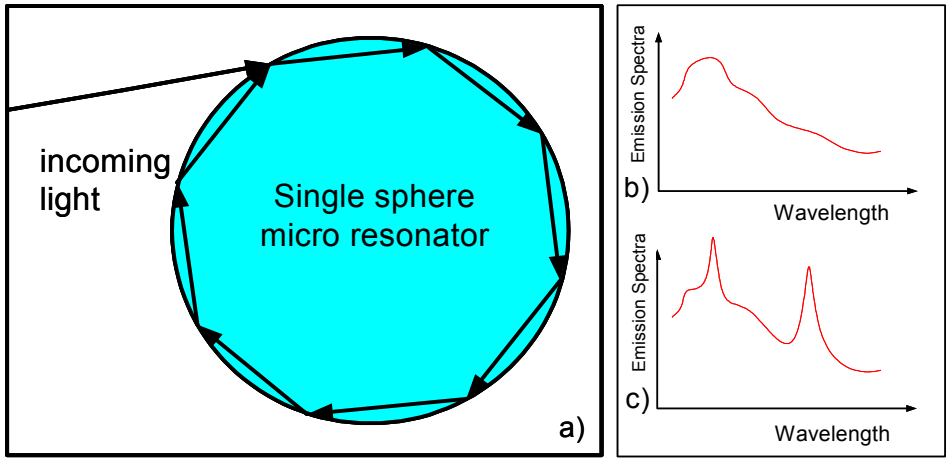

Figure 1.

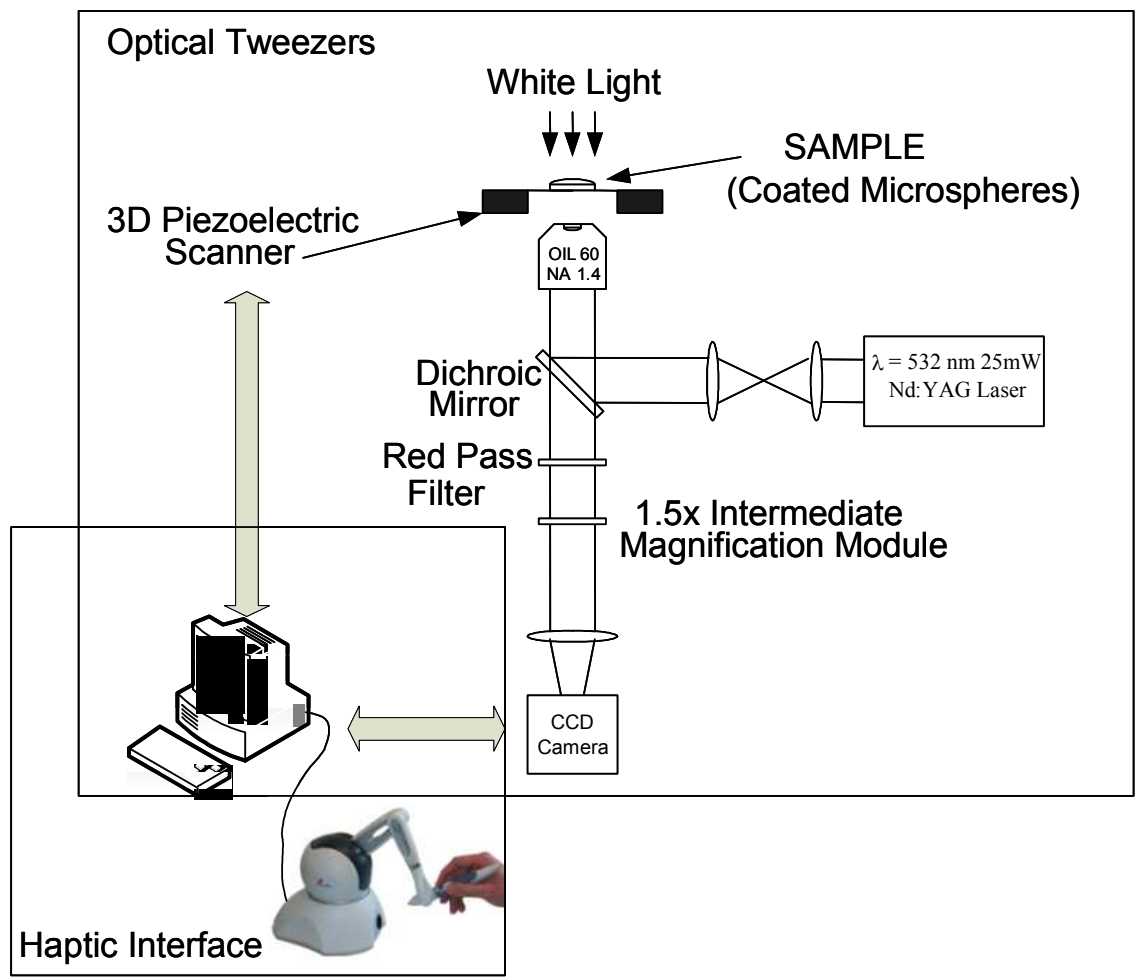

Figure 2.

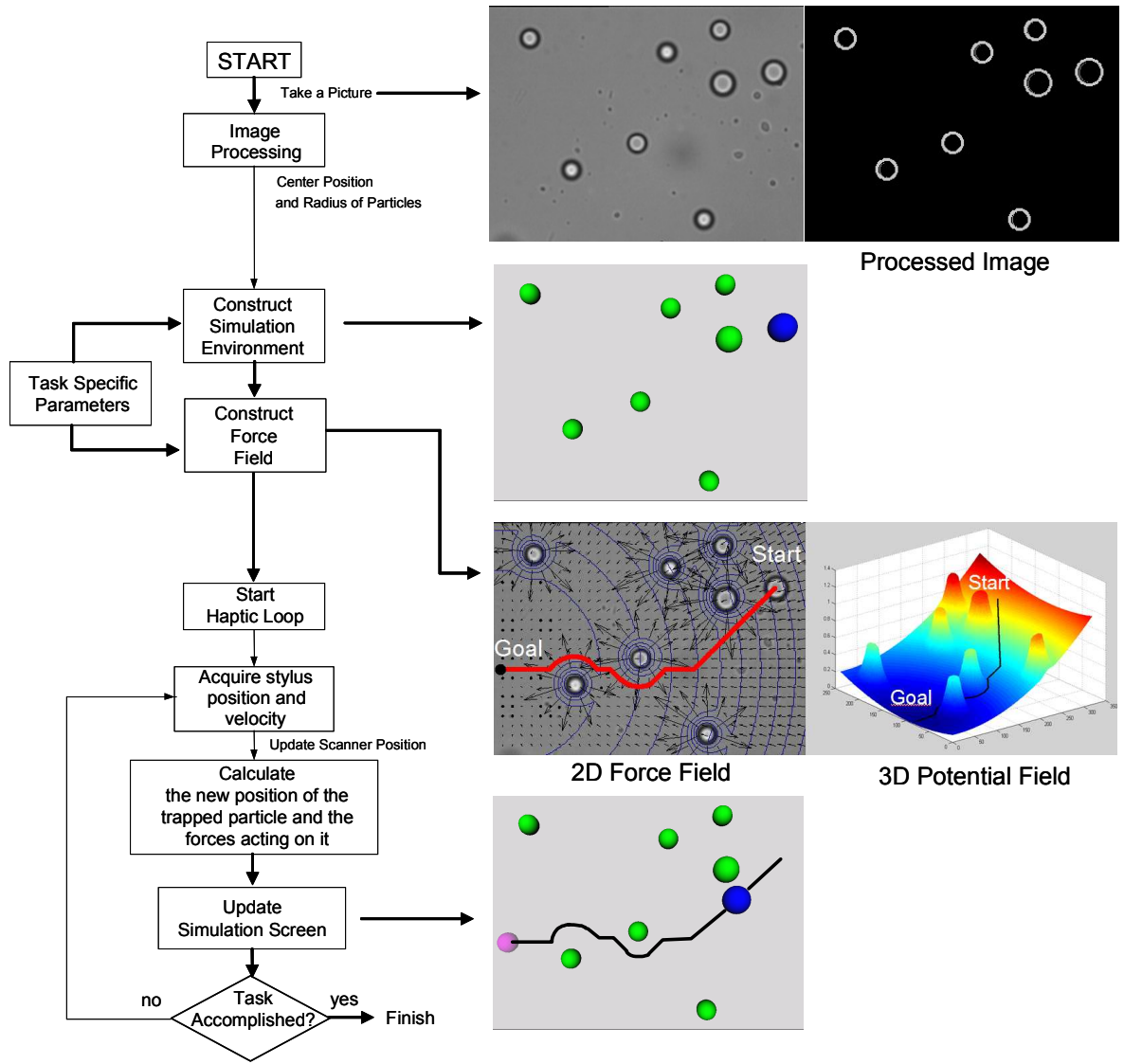

Figure 3.

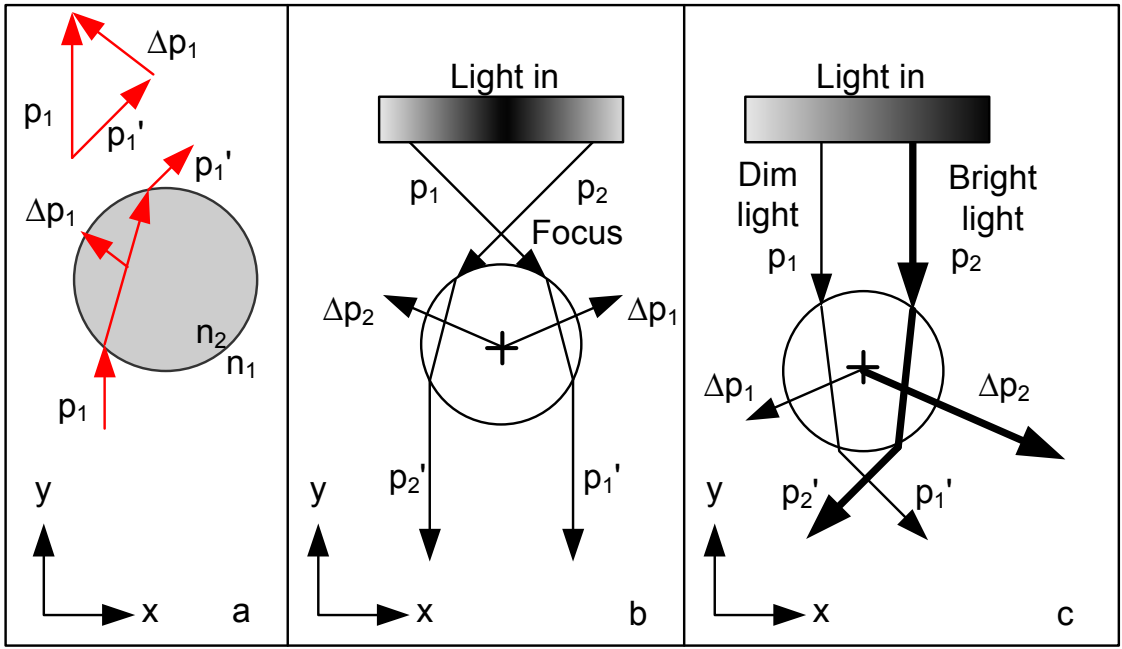

Figure 4.

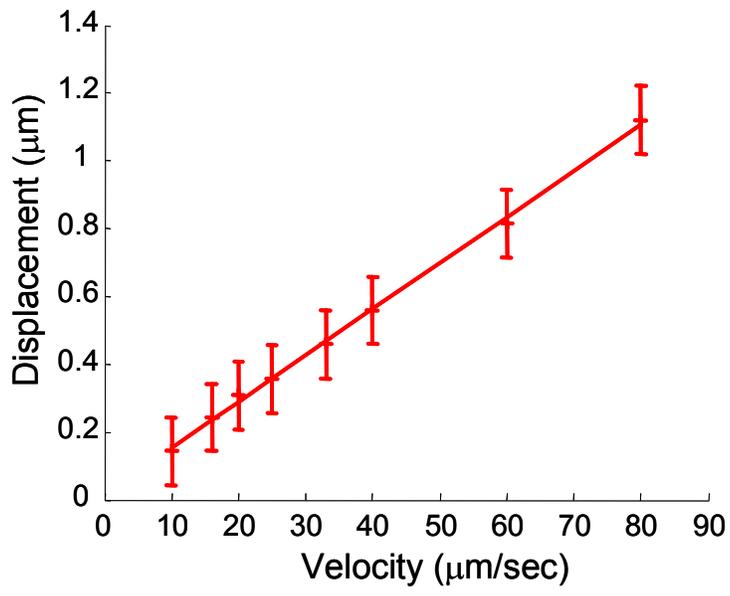

Figure 5.

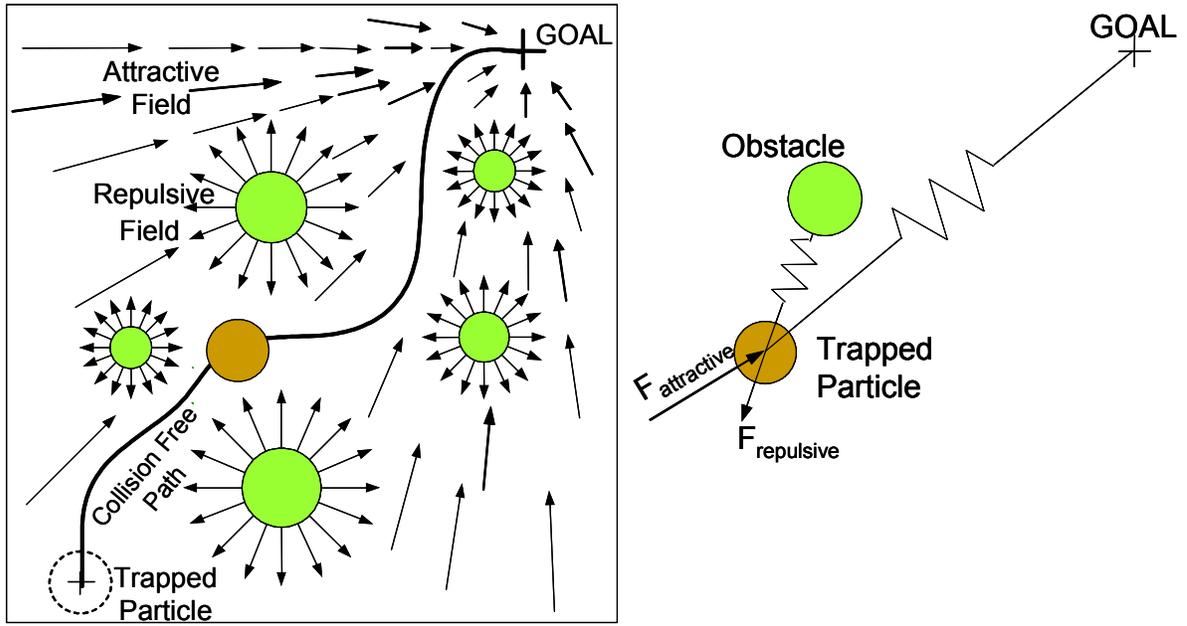

Figure 6.

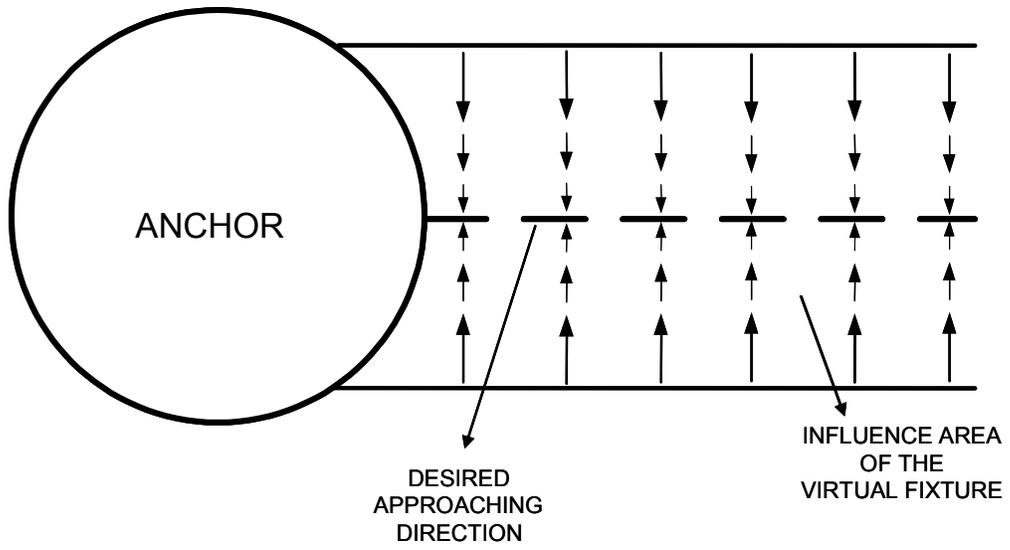

Figure 7.

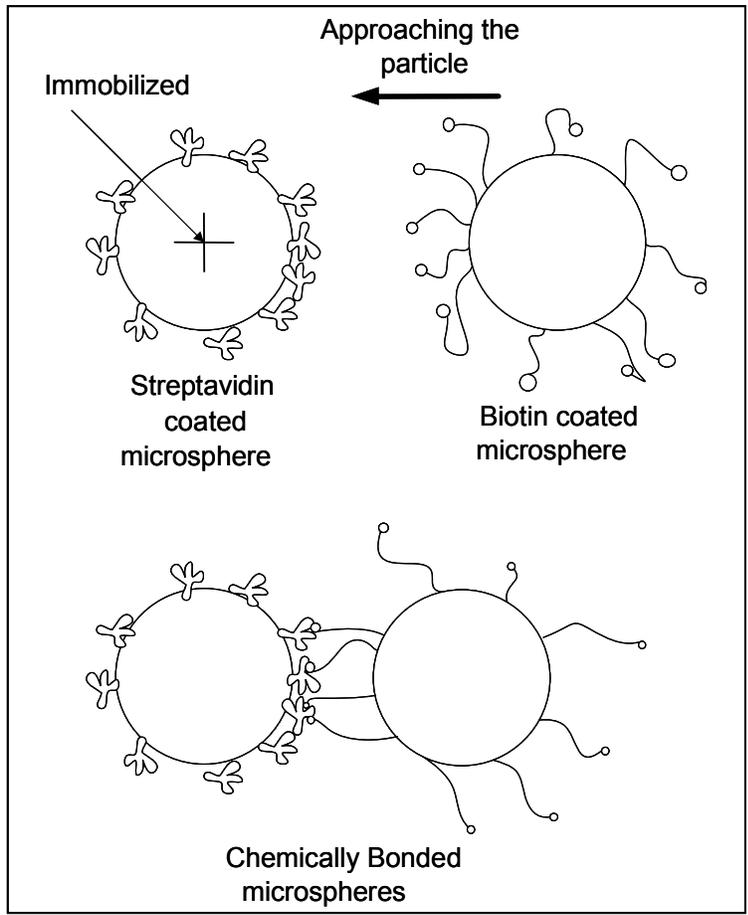

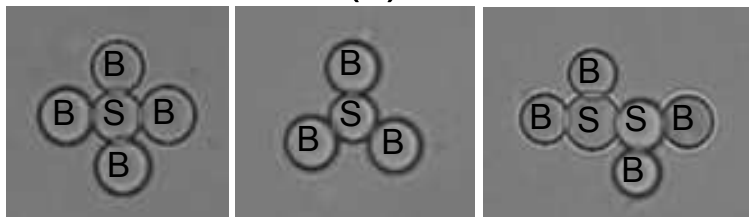

Figure 8.

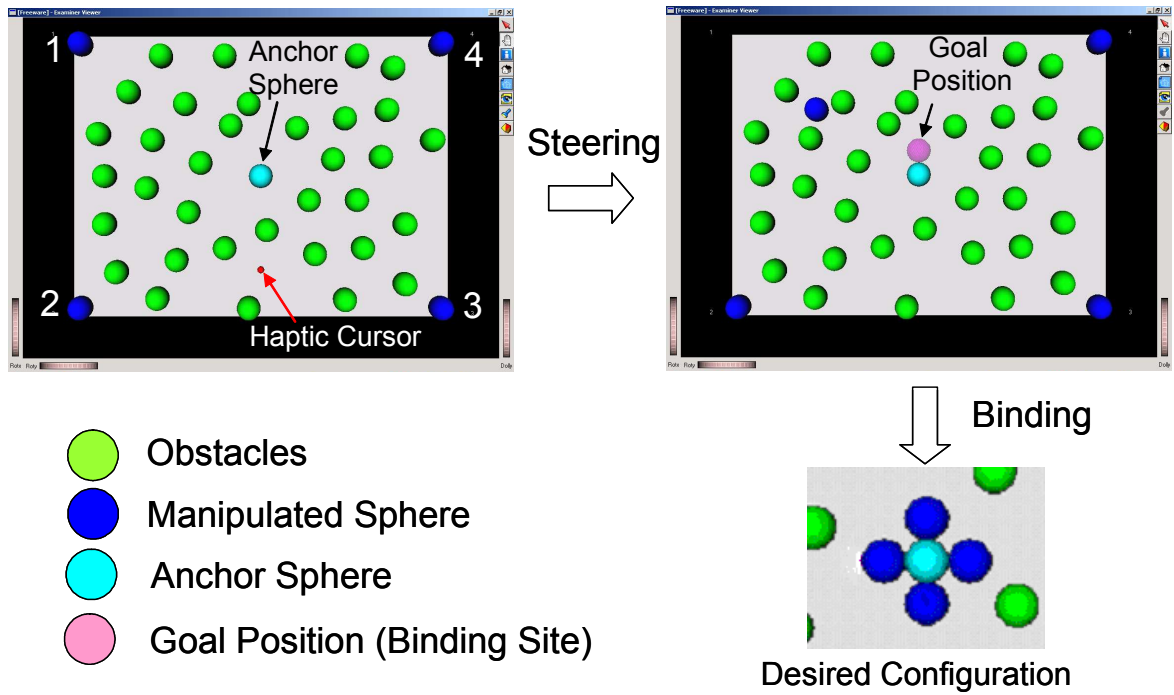

Figure 9.

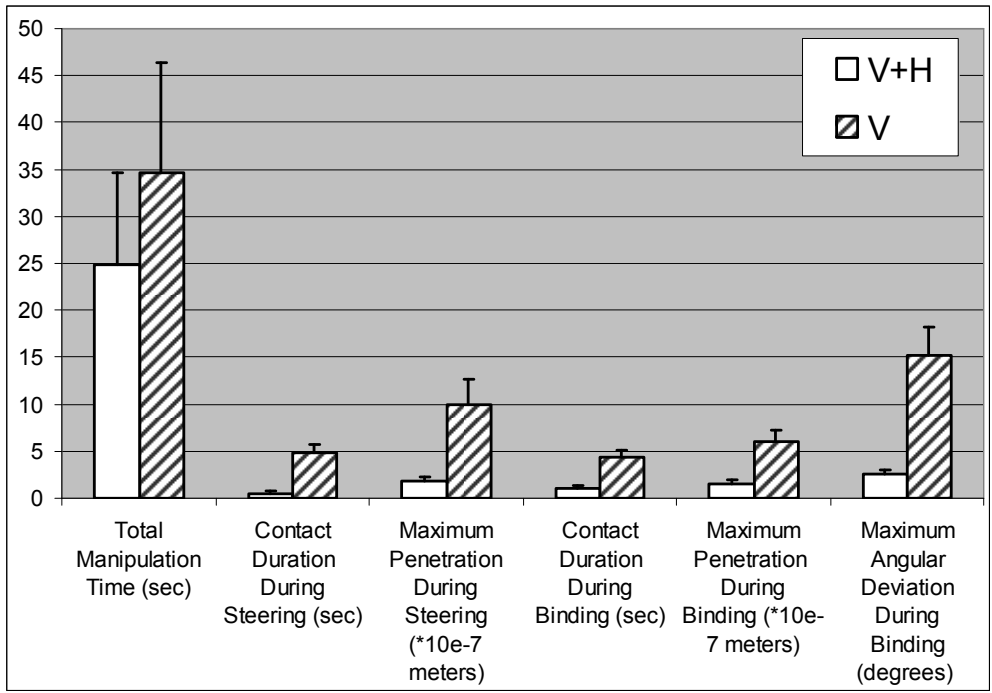

Figure 10.

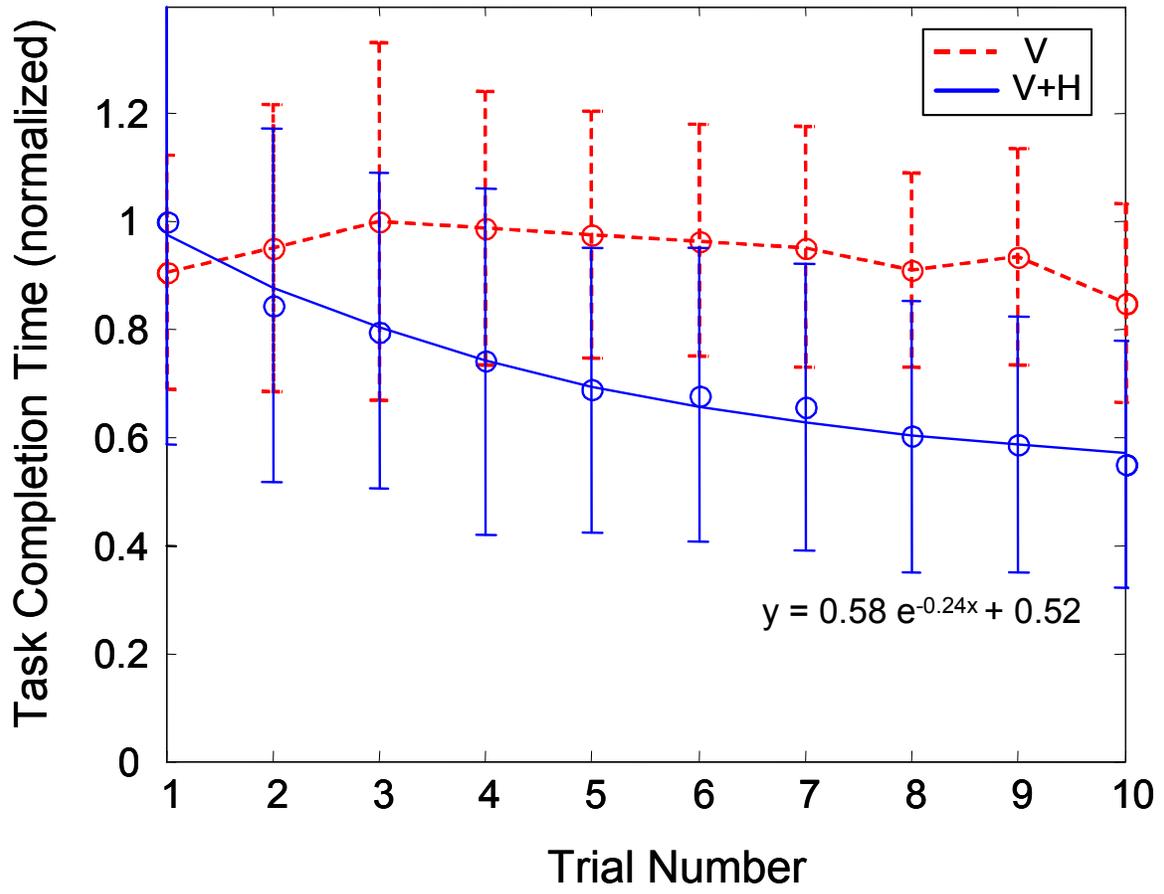

$y = 0.58\, e^{-0.24x} + 0.52$

Figure 11.

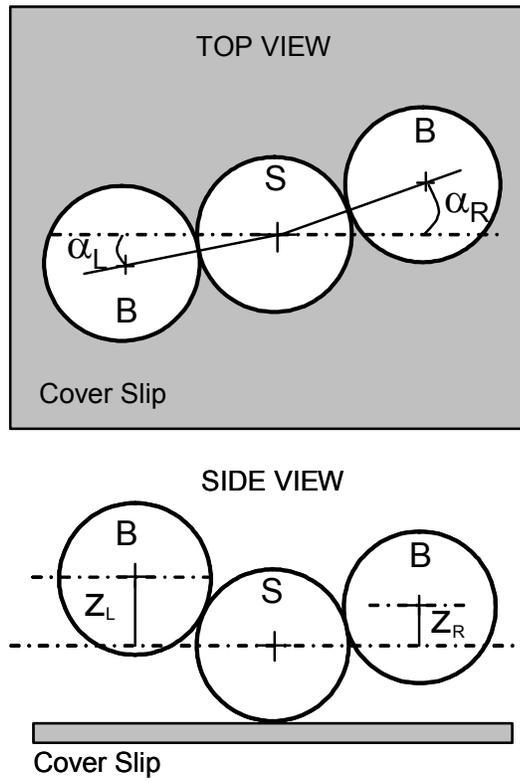

Figure 12.

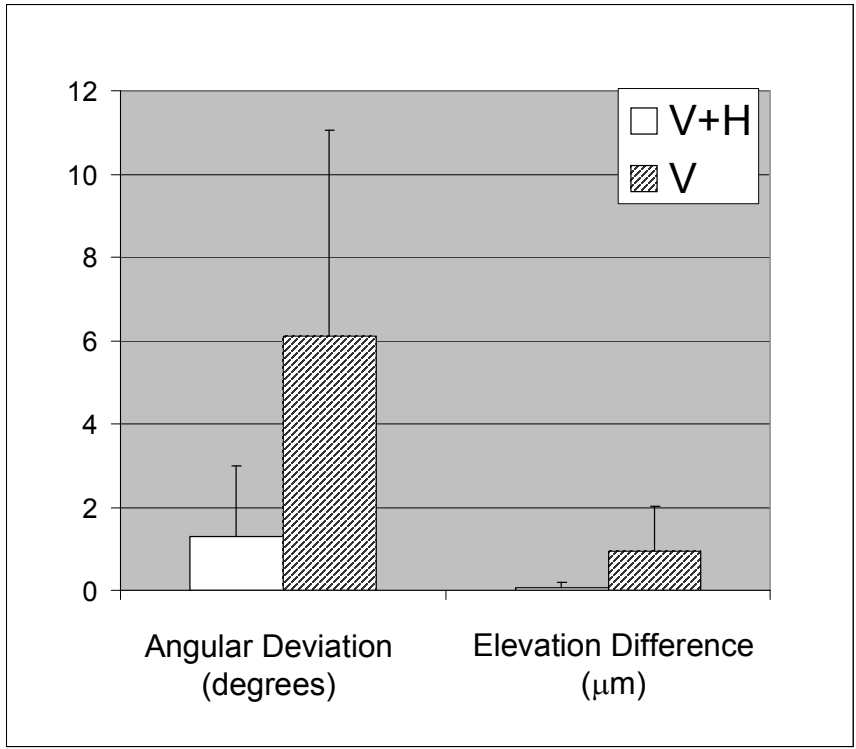

Figure 13.

**Figure Captions:**

**Figure 1.** a) Whispering Gallery Modes (WGMs): At these resonant wavelengths, the light undergoes total internal reflection inside the sphere and folds back on itself in phase. The emission spectra of an atom before (b) and after (c) entering an optical cavity (representative). The resonant modes at two different wavelengths are apparent in (c).

**Figure 2.** The experimental set-up for manipulating microspheres include an OT and a haptic device. The haptic device controls the movements of the 3D piezo scanner with respect to the laser trap and displays drag and artificial guidance forces to the user during the real-time manipulations.

**Figure 3.** The flow-chart of the software interface for manipulating the micro spheres with the guidance of a haptic device. First, the snapshot taken by a CCD camera is processed to identify the size and the locations of the microspheres. An identification number (ID) is assigned to each particle in the scene. Then, the simulation environment is constructed with this information in hand and based on the task specific parameters entered by the user. The user enters the IDs of the trapped and anchor spheres and the geometry of the desired pattern to be constructed. The system displays the ideal binding sites and angles in the simulation window according to the user input. Then, the artificial force field is created automatically and the haptic loop is started. As the user moves the haptic stylus, the piezo scanner moves the spheres on the coverslip and the forces acting on the particle are displayed to the user via the haptic device.

**Figure 4.** For the ray optics regime (the diameter of the trapped particle is much larger than the wavelength of the incident light), a single ray can be tracked through the particle. If the ratio of the refractive index of the particle to the refractive index of the medium surrounding it is sufficiently large, then diffraction effects can be neglected. In order to satisfy this condition, particles to be manipulated are put into an aqueous solution such as water. The incident laser beam can be decomposed into individual rays according to their intensity (at the center of the beam the most intense light is present. A logarithmic decay occurs as the distance from the center is increased). After a ray travels through the particle, the momentum of the photons are changed and the force due to this change applies optical forces on the particle which has components in both the forward (scattering force) and side (gradient force) directions (see the X and Y components of $\Delta P_1$ in Figure a). As long as the particle is at the center (Figure b), the intensity of the light around the particle is symmetric and the gradient forces (see X components of $\Delta P_1$ and $\Delta P_2$) cancel each other resulting in a forward force (the summation of Y components of $\Delta P_1$ and $\Delta P_2$) which is balanced by the weight of the particle. However, when the particle is eccentric according to the beam, the net force on the particle creates a gradient force which always pulls the particle to the center of the beam (Figure c). Hence, as the laser beam is moved, the particle, under the effect of this gradient force, moves with the laser. Or, the laser beam trapping a particle is stationary, but a piezo scanner moves the other particles with respect to the trapped particle as in our case.

**Figure 5.** The OT must be calibrated to calculate the maximum manipulation speed. For calibration, the piezo scanner is commanded to move at increasing speeds until the particle escapes from the trap while the distance between the centers of the laser beam and the trapped particle is recorded.

**Figure 6.** During the steering of a trapped particle, guidance forces due to an artificial potential field and the drag force acting on the particle are displayed to the user through the haptic device (a). The manipulated particle is pulled to the goal point via the attractive forces in the field while the repulsive forces around the other particles along the steering path prevent the undesired collisions (b).

**Figure 7.** During the binding of a biotin-coated particle to a streptavidin coated one (the anchor), the virtual fixture helps the user to maintain the necessary angle of approach. The guidance forces that are linearly decreasing and pushing the biotin particle towards the line of approach are displayed to the user through the haptic device. Also, the repulsive forces around the streptavidin particle due to the potential field prevent the user to apply excessive forces during the binding.

**Figure 8.** a) Assembly of microspheres using biotin-streptavidin binding. If a biotin coated sphere is brought to contact with a streptavidin coated sphere, the binding of the biotin to the suitable docking sides on the streptavidin occurs rapidly as a result of the strong affinity between those two. b) Patterns constructed using our set-up. "B" and "S" letters indicate the biotin and streptavidin-coated spheres respectively.

**Figure 9.** The simulation window displayed to the subjects during Experiment I. The task is to steer each of the four blue spheres located at the corners of the cover slip (gray plane) and bind it to the anchor sphere (light blue sphere) located at the center to construct the desired configuration.

**Figure 10.** The results of the Experiment I conducted in virtual environments show that manipulating micro particles with haptic and visual feedback together (V+H) is significantly better than that of visual feedback only (V) in all measures of performance.

**Figure 11.** The effect of haptic feedback on the task learning in Experiment I. The task completion times are plotted against the number of trials. When only visual feedback is displayed to the subjects, learning did not occur (dashed lines). However, when the haptic feedback is added to the visual information, the task completion time shows an exponential decay indicating that the task learning has occurred.

**Figure 12.** The performance measures used in Experiment II: At the end of the task, the image of the pattern constructed by the subject is processed to measure the deviations from the desired angle of binding ($\alpha_L$ and $\alpha_R$) on the plane of the coverslip. Moreover, the positional deviations from the desired binding position along the axis perpendicular to the coverslip (i.e. parallel to the axis of the laser beam) are measured using the camera focus and the recorded movements of the scanner.

**Figure 13.** The results of the Experiment II conducted with the experimental set-up show that haptic and visual feedback together (V+H) is significantly better than visual feedback only (V) in assembling spheres to form desired patterns.

**Tables:**

|         | GROUP 1 (4)                  | GROUP 2 (4)                  |
|---------|------------------------------|------------------------------|
| SET 1   | V+H (10) → V(10)             | V (10) → V+H(10)             |
|         | ↓ 1 WEEK                     | ↓ 1 WEEK                     |
| SET 2   | V (10) → V+H(10)             | V+H (10) → V(10)             |

Table 1.

| Subject | V<br>Number of Trials | V+H<br>Number of Trials |
| --- | --- | --- |
| 1 | 5/7 | 5/5 |
| 2 | 5/6 | 5/5 |
| 3 | 5/7 | 5/5 |
| 4 | 5/5 | 5/5 |
| 5 | 5/9 | 5/6 |
| 6 | 5/6 | 5/5 |
| 7 | 5/6 | 5/5 |
| 8 | 5/8 | 5/5 |
| **Total** | **40/54 (74%)** | **40/41 (97%)** |

Table 2.

**Table Captions:**

**Table 1.** The matrix design for Experiment I. Subjects were divided into 2 groups (Group I and Group II), each having 4 members. The experiment was performed in two sets with one week rest interval between them. Each set consists of 20 trials. In set I, Group I performed the experiment 10 times under the condition (V) first and then condition (V+H) while Group II performed the experiment 10 times under the condition (V+H) first and then condition (V). In set II (one week later), groups repeated the same experiment in exactly the opposite order of sensory conditions.

**Table 2.** The success rates of the subjects in assembling 3 spheres on a horizontal line under the conditions (V) and (V+H) in Experiment II.